\documentclass[useAMS,usenatbib,usegraphicx]{mn2e}
\def\spose#1{\hbox to 0pt{#1\hss}}
\def\approxlt{\mathrel{\spose{\lower 3pt\hbox{$\sim$}}
        \raise 2.0pt\hbox{$$<$$}}}
\def\approxgt{\mathrel{\spose{\lower 3pt\hbox{$\sim$}}
        \raise 2.0pt\hbox{$>$}}}
\newcommand{\gta}{\mathrel{\spose{\lower 3pt\hbox{$\mathchar"218$}}
      \raise 2.0pt\hbox{$\mathchar"13E$}}}

\title[Formation of Black Holes by Direct Collapse]{Formation of Supermassive Black Holes by Direct Collapse in Pregalactic Halos}

\author[M. C. Begelman, M. Volonteri \& M.J. Rees] {Mitchell C. Begelman
$^{1, 2, 3}$\thanks{E-mail: mitch@jila.colorado.edu (MB); marta@ast.cam.ac.uk (MV); mjr@ast.cam.ac.uk (MR)}, Marta Volonteri$^{3\star}$, and Martin J. Rees$^{3}$\footnotemark[1]\\ $^{1}$ JILA, University of
Colorado, Boulder, CO 80309-0440, USA \\ $^{2}$ Also at Department of Astrophysical and Planetary Sciences, University of Colorado at Boulder  \\ $^{3}$ Institute of Astronomy, Madingley Road, Cambridge CB3 0HA, UK }

\begin{document}
\maketitle

\begin{abstract}
We describe a mechanism by which supermassive black holes can form directly in the nuclei of protogalaxies, without the need for ``seed" black holes left over from early star formation. Self-gravitating gas in dark matter halos can lose angular momentum rapidly via runaway, global dynamical instabilities, the so-called ``bars within bars" mechanism. This leads to the rapid buildup of a dense, self-gravitating core supported by gas pressure --- surrounded by a radiation pressure-dominated envelope --- which gradually contracts and is compressed further by subsequent infall.  We show that these conditions lead to such high temperatures in the central region that the gas cools catastrophically by thermal neutrino emission, leading to the formation and rapid growth of a central black hole.  

We estimate the initial mass and growth rate of the black hole for typical conditions in metal-free halos with $T_{\rm vir} \sim 10^4$ K, which are the most likely to be susceptible to runaway infall.  The initial black hole should have a mass of $\la 20 M_\odot$, but in principle could grow at a super-Eddington rate until it reaches $\sim 10^4-10^6 M_\odot$.  Rapid growth may be limited by feedback from the accretion process and/or disruption of the mass supply by star formation or halo mergers.  Even if super-Eddington growth stops at $\sim 10^3-10^4 M_\odot$, this process would give black holes ample time to attain quasar-size masses by a redshift of 6, and could also provide the seeds for all supermassive black holes seen in the present universe.   

\end{abstract}


\section{Introduction}

Several scenarios have been presented for the formation and growth of supermassive black holes (SMBHs) in the nuclei of galaxies. One possible route traces the black-hole progenitors back to the first generation of stars. The first stars formed out of metal-free gas, with the lack of an efficient cooling mechanism possibly leading to a very top-heavy initial stellar mass function (Larson 1998; Carr, Bond \& Arnett 1984). Numerical simulations of the fragmentation of primordial clouds in standard cold dark matter theories suggest that Pop III stars were indeed very massive (Bromm, Coppi \& Larson 1999, 2002; Abel, Bryan \& Norman 2000), and would have left behind black hole ``seeds" of anywhere from 10's to several hundred solar masses.  The main features of a plausible scenario for the hierarchical assembly, growth, and dynamics of massive black holes from such seeds have been discussed most recently by Volonteri, Haardt \& Madau (2003), Volonteri et al. (2005), and Volonteri \& Rees (2005).  

Another family of models for massive black hole formation is based on the collapse of supermassive objects formed directly out of dense gas (Haehnelt \& Rees 1993; Umemura, Loeb \& Turner 1993; Loeb \& Rasio 1994; Eisenstein \& Loeb 1995; Bromm \& Loeb 2003; Koushiappas, Bullock \& Dekel 2004).  The main challenge for these models is the disposal of angular momentum.  Eisenstein \& Loeb (1995) and Koushiappas et al. (2004) investigated the formation of black holes from low angular momentum material, either in halos with extremely low angular momentum (Eisenstein \& Loeb 1995), or by considering only the low angular momentum tail of material in halos with efficient gas cooling.  But even in these models, as in all the others, substantial angular momentum transport is required in order for the gas to form a central massive object, which ultimately collapses  as a result of the post-Newtonian gravitational instability.  Various angular momentum transport mechanisms have been invoked, including radiation drag against the cosmic microwave background (at very high redshifts: Umemura et al. 1993), viscosity driven by magnetic fields or turbulence, Rossby waves (Colgate et al. 2003) and self-gravitational instabilities.

The scenario we investigate here is related to the second family of models, and focuses on the outcome of global dynamical instabilities driven by self-gravity, the so-called ``bars within bars" mechanism (Shlosman, Frank \& Begelman 1989; Shlosman, Begelman \& Frank 1990).  Self-gravitating gas clouds become bar-unstable when the level of rotational support surpasses a certain threshold. A bar can transport angular momentum outward on a dynamical timescale via gravitational and hydrodynamical torques, allowing the radius to shrink.  Provided that the gas is able to cool, this shrinkage leads to even greater instability, on shorter timescales, and the process cascades.  This mechanism is a very attractive candidate for collecting gas in the centers of halos, because it works on a dynamical time and can operate over many decades of radius. In contrast to the formation of a supermassive ``star", with high entropy throughout, we show that the ``bars within bars" mechanism produces a ``quasistar" with a very low specific entropy near the center.  As a result, the initial core collapse leading to black hole formation involves only a few solar masses, rather than the thousands of solar masses usually associated with direct collapse models.  Despite this modest beginning, accretion from the envelope surrounding the collapsed core can build up a substantial black hole mass very rapidly --- possibly at a highly super-Eddington rate.  

The plan of the paper is as follows. In \S~2, we discuss the criterion for global gravitational instability and apply it to the gas in dark matter halos with a realistic distribution of angular momentum parameters.   If more than a few tenths of the baryonic matter fall toward the center of the halo, then gravitational instability should be very common.  But even an infall fraction of $\sim 10$\% can lead to an interesting number of unstable halos. In \S~3 we specialize to halos with virial temperatures $T_{\rm vir} \ga 10^4$ K and metal-free gas. The bars within bars scenario makes specific predictions about the radial distribution of infalling gas and the associated circular velocity, which goes from constant in the outer parts of the inflow to quasi-Keplerian close-in.  We first discuss the infall processes neglecting star formation, and then show how the process is modified (but not necessarily halted) if a fraction of the infalling gas at each radius forms stars.  The gravitational binding energy liberated by infalling gas increases steadily with decreasing radius, until the luminosity exceeds the Eddington limit, the infall stalls, and a radiation pressure-supported ``quasistar" forms (\S~4).  The radius of the quasistar is a few astronomical units, a scale that does not change even as the quasistar grows in mass. 

We show that the quasistar has a positive specific entropy gradient, and that gas pressure remains important in the quasistar's core.  The temperature of this core steadily increases as matter piles on to the quasistar (\S~5), eventually approaching $10^9$ K, at which point it undergoes catastrophic cooling and collapse by thermal neutrino emission (\S~6).  We argue that this leads to the formation of a black hole of $\sim 10-20 M_\odot$, which may then grow at rate that greatly exceeds the Eddington limit (\S~7). This rapid growth could produce a black hole of several million solar masses, although feedback and depletion of the mass supply could quench the growth rate at an earlier stage. We discuss the co-evolution of the black holes and their hosts and the global impact of the black hole population in \S~8. We conclude by discussing the implications of this model for the interpretation of high-$z$ quasars, the statistics of black hole masses in the local universe, and its relevance to other astrophysical situations where black holes could grow at a very rapid rate.  

Unless otherwise stated, all results shown below refer to the currently favored $\Lambda$CDM world model with $\Omega_M=0.3$, $\Omega_\Lambda=0.7$, $h=0.7$, $\Omega_b=0.045$, $\sigma_8=0.93$, and $n=1$.

\section{Conditions for runaway collapse}

We focus here mainly on the dynamical stability of the gas in halos with virial temperatures $T_{\rm vir} \ga 10^4$K.  Runaway collapse could also occur in smaller halos, provided that molecular hydrogen cooling is efficient and gas can cool well below the virial temperature.  In the absence of molecular hydrogen, gas in halos with $T_{\rm vir} < 10^4$ K would tend to remain less dense than the dark matter; tidal forces would then prevent widespread collapse and fragmentation at this stage. Since cooling and collapse of the gas is more likely in large halos, and the masses involved are larger, we henceforth refer to halos with virial temperatures $T_{\rm vir} \ga 10^4$K , unless otherwise stated. We stress nevertheless that runaway collapse is not completely ruled out in smaller systems at early times, well before the first generation of stars created a photodissociating background. 

Bromm \& Loeb (2003) show that if molecular hydrogen formation is suppressed in halos with $T_{\rm vir}> 10^4$ K, the gas tends to condense into massive clumps in the center. The gaseous component of these halos can cool even in the absence of ${\rm H_2}$ via neutral hydrogen atomic lines to $\sim 8000$ K, and contract nearly isothermally (Oh \& Haiman 2002).  These massive clumps do not fragment as long as molecular hydrogen remains unimportant. One way to hinder the formation of molecular hydrogen is the presence of a dissociating background (Haiman, Abel \& Rees 2000; Bromm \& Loeb 2003; Oh \& Haiman 2002; but see Machacek, Bryan \& Abel 2001).  It is therefore possible that the formation of seed black holes in massive halos follows an earlier epoch of star formation.  We consider redshifts high enough that a large fraction of gas is still unenriched by metals, or very lightly polluted, so that metal line cooling is still unimportant (Santoro \& Shull 2006). 

We assume that the baryons preserve their specific angular momentum during collapse (Mo, Mao \& White 1998), and settle into a rotationally supported disc at the center of the halo (Mo et al. 1998; Oh \& Haiman 2002).
Flattened systems can be subject to dynamical and secular instabilities, even when
embedded in external halos (Fall \& Efstathiou 1980). Several instability 
criteria have been investigated (e.g., Ostriker \& Peebles 1973; Efstathiou, Lake 
\& Negroponte 1982; Christoudoulou, Shlosman \& Tohline 1995), which determine the maximum 
rotational energy (or angular momentum) that a system can possess and still be stable against 
bar-like instabilities. Christoudoulou et al. (1995) propose a simple, but robust,
criterion for stability which can be expressed as 
\begin{equation}
\alpha=\left(\frac{1}{2}f\frac{T}{|W|}\right)^{1/2}<0.34,
\label{instcrit}
\end{equation}
where $T$ is the rotational kinetic energy, $W$ is the gravitational potential energy, 
and $f$ is a parameter dependent on the geometry of the system, with $f=1$ for discs. 

Numerical simulations have not yet clarified the detailed dynamics of gaseous collapse in halos, and we explore here three different models for self-gravitating gas discs.  We assume that the disc has either constant circular velocity (Mestel discs: Mestel 1963) or constant angular velocity (rigid body rotation), or that the gas settles down into a classical exponential disc. We embed the gaseous discs into a halo of mass $M_h$, virial radius $R_{\rm vir}$, and  virial 
temperature $T_{\rm vir}$, described by a Navarro, Frenk \& White (1997, hereafter NFW) dark matter density
profile, with a  spin parameter $\lambda_{\rm spin}$ ($\equiv J_h E_h^{1/2}/ GM_h^{5/2}$, where $J_h$ is the total angular momentum and $E_h$ is the binding energy).  We recall that, within the spherical collapse model, the mass of a halo, at a given redshift of formation, 
scales with the virial temperature as
$M_h\simeq 10^4 \Delta_{\rm vir}^{-1/2} T^{-3/2}_{\rm vir} M_\odot$, where $\Delta_{\rm vir}$ is the virial density in units of the 
critical density.

We determine the characteristics of the gaseous discs via a procedure similar to that of Mo et al. (1998). We then apply the Christoudoulou et al. (1995) criterion in order to determine the stability of the modeled systems.  Stability depends on two parameters, the halo spin parameter $\lambda_{\rm spin}$, and the fraction of baryonic matter that ends up in the disc, $f_d = (\Omega_M/\Omega_b) (M_{\rm disc}/M_h)$.  The results  can be understood qualitatively by approximating the disc kinetic energy as 
$T_{\rm disc}\approx 0.5\,M_{\rm disc}\, V_c^2(R_{\rm disc})$, where $R_{\rm disc}$ is the scale length of the disc,\footnote{Note that both the Mestel disc and the rigid disc are defined only for $R<R_{\rm disc}$ (Mestel 1963).}
which can be determined under the assumption that the collapsing baryons conserve angular momentum. 
If we ignore the contribution of the halo to the circular velocity, for the three cases we find:
\begin{equation}
T_{\rm disc,Mestel}\approx 
\frac{\pi^2\,G\,M_{\rm disc}^2f_d (\Omega_b/\Omega_M)}{32\,\lambda_{\rm spin}^2\,R_{\rm vir}};
\end{equation}
\begin{equation}
T_{\rm disc,rigid}\approx 
\frac{9\,\pi^2\,G\,M_{\rm disc}^2f_d (\Omega_b/\Omega_M)}{40\,\lambda_{\rm spin}^2\,R_{\rm vir}};
\end{equation}
\begin{equation}
T_{\rm disc,exp}\approx 
\frac{G\,M_{\rm disc}^2}{\lambda_{\rm spin}\,R_{\rm vir}\,f_R};
\end{equation}
where 
\begin{equation}
f_R \approx [1-3 f_d (\Omega_b/\Omega_M) +5.2 f_d^2 (\Omega_b/\Omega_M)^2 ]
\end{equation}
(see Mo et al. 1998 for the exact expression). 
At fixed $\lambda_{\rm spin}$ and $f_{d}$, $T_{\rm disc,rigid}>T_{\rm disc,Mestel}>T_{\rm disc,exp}$. 
The total kinetic ($T$) and potential ($|W|$) energies of the systems, including the contribution and
stabilizing effect of the NFW halo,   
increase with $\lambda_{\rm spin}$ due to the halo contribution within $R_{\rm disc}$, which increases with $\lambda_{\rm spin}$. 
The ratio $T/|W|$, nevertheless, decreases, due to an increasingly dominant halo contribution. 

\begin{figure}
\includegraphics[width=8cm, height=8cm]{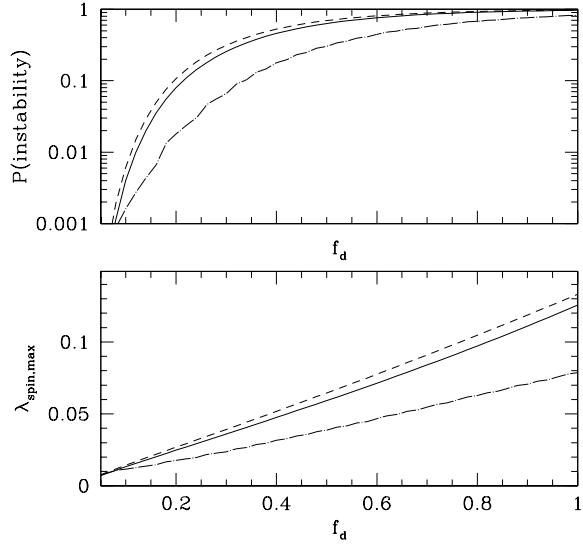}
\caption{Bottom panel: maximum spin parameter, $\lambda_{\rm spin,max}$, for disc instability as a function of 
the gas fraction ending up in the disc. Discs are stable for $\lambda_{\rm spin}>\lambda_{\rm spin,max}$. 
Solid line: Mestel disc, dashed line: rigid disc, dot-dashed line: exponential disc. Upper panel: fraction of unstable discs, for each of the three disc models. }
\label{Pinst}
\end{figure}

The stability results are summarized in Fig.~\ref{Pinst}, in which the maximum spin parameter 
$\lambda_{\rm spin,max}$ for which a disc is unstable is shown as a function of the fraction of baryons forming the disc, 
i.e., for every $f_d$, discs are stable for $\lambda_{\rm spin}>\lambda_{\rm spin,max}$. 

The distribution of spin parameters found in numerical simulations is well fit by a lognormal distribution in $\lambda_{\rm spin}$, with mean 
$\bar \lambda_{\rm spin}=0.05$ and standard deviation $\sigma_\lambda=0.5$:
\begin{equation} 
\label{plambda}
p(\lambda) \,d\lambda
={1\over \sqrt{2\pi} \sigma_\lambda}
\exp \left[-{\ln ^2 (\lambda/{\bar \lambda})
\over 2 \sigma_\lambda^2}\right] {d\lambda \over \lambda},
\end{equation}
This function is a good fit to the N-body results of Warren et al (1992). Similar results were found in later investigations (e.g., 
Cole \& Lacey 1996; Bullock et al. 2001; van den Bosch et al. 2002). With this assumption we can estimate the fraction of discs subject to dynamical instability, as a function of $f_d$, for each of the three disc models (Fig.~\ref{Pinst}).

\section{Structure and evolution of collapsing gas}

The unstable conditions described in \S~2 are expected to lead to runaway infall, provided that the gas remains cooler than the local virial temperature as it collapses.  For the densities expected in pregalactic haloes, the cooling time to $\la 10^4$ K is much smaller than the dynamical time; this ordering is preserved as the collapse proceeds.  At the initial disk radius, the gravitational potential due to the gas is already appreciable compared to that of the dark matter.  As we will see below, collapse leads to a mean gas density profile at least as steep as $r^{-2}$, implying a virial temperature that remains roughly constant or increases with decreasing $r$, whereas the dark matter density is expected to increase only as $\propto r^{-1}$ in the inner parts of the halo (NFW).  The relative dominance of gas self-gravity over the dark matter potential thus increases as the gas collects toward the center, implying that the conditions for large-scale gravitational instability intensify with decreasing radius.  Conditions are therefore ideal for the ``bars within bars" instability.  We will henceforth neglect the dark matter.  

The collapse of a self-gravitating, isothermal gas cloud has been analyzed in both the nonrotating (Larson 1969; Penston 1969; Shu 1977) and rotating but inviscid (Saigo \& Hanawa 1998; and references therein) limits. In all cases, the outer part of the flow evolves toward the density profile of a singular isothermal sphere, $\rho \propto r^{-2}$.  Mineshige \& Umemura (1997), analyzing the case of a shrinking, self-gravitating accretion disk subject to an $\alpha-$viscosity, found analogous behavior, i.e., a surface density distribution $\Sigma \propto r^{-1}$ in the outer regions. Using SPH simulations, Englmaier \& Shlosman (2004) show that a rotating, self-gravitating gas cloud can decouple dynamically from a larger stellar bar, thus verifying the basic bars-within-bars picture.  They find that the shrinking gas bar develops a strong density gradient with radius, and an angular pattern speed $\Omega \propto a^{-1}$, where $a$ is the semimajor axis of the gaseous bar.  These features are also consistent with an $r^{-2}$ radial density distribution.     

The $\rho \propto r^{-2}$ behavior can be understood as follows.  Suppose the initial mass and radius of the cloud are $M_0$ and $r_0$, respectively, corresponding to a virial speed $v_0 = (GM_0/ r_0)^{1/2} \sim 10 v_{10}$ km s$^{-1}$.  At large radii and early times, the evolutionary timescale of the flow is set by the free fall time at $r_0$, $t_0 \sim r_0/v_0$.  At smaller radii the free fall time is shorter, but matter is being fed into these radii on the much longer timescale $t_0$.  Therefore, the mass flux through these regions must be roughly constant and independent of radius, 
\begin{equation}
\label{mdotouter}
\dot M(r)= \dot M_0 \sim {M_0\over t_0} \sim {v_0^3 \over G} = 0.2\ v_{10}^3 \ M_\odot \ {\rm yr}^{-1} .  
\end{equation}
On average, the self-gravitating gas at every radius gets rid of its angular momentum on the local free fall time, $t_{\rm ff}(r) \sim [GM(r)/ r^3]^{-1/2}$, where $M(r)$ is the mass contained within a radius $r$.  The mass flow rate is therefore $\dot M_0 \sim M(r)/t_{\rm ff}(r) \propto [M(r)/r]^{3/2}$. Since $\dot M_0$ is constant, we have $M(r)\propto r$ and $\rho \propto r^{-2}$.   

This behavior can change if the collapsing gas fragments and forms stars.  A self-gravitating cluster of collisionless particles would increase its velocity dispersion in response to the bar potential, thus quenching the instability.  Although the cooling timescale of the gas is shorter than the dynamical time, and therefore violates the Gammie (2001) criterion for avoiding fragmentation, we do not expect fragmentation to deplete a large fraction of the inflow.  This is because the circular speed in the potential of the self-gravitating gas remains roughly constant at $r < r_0$, and corresponds to a temperature that is close to the virial temperature of the halo.  Since we consider only halos with $T_{\rm vir} \ga 10^4$ K, the gas never cools very far below $T_{\rm vir}$ and therefore does not form a very thin sheet (i.e., the Toomre [1964] parameter $Q$ does not become extremely small).  The Jeans mass under such conditions is only a few times smaller than $M(r)$, suggesting that fragmentation will be inefficient.  A corollary of this argument is that the efficient collection of gas in the center of a halo might occur only for a relatively narrow range of $T_{\rm vir} \ga 10^4$ K, and only under metal-free conditions where the formation of H$_2$ is inhibited (these arguments were discussed at length by Bromm \& Loeb 2003), or for gas enriched below the critical metallicity threshold for fragmentation (Santoro \& Shull 2006). 

As noted in many earlier works, the $r^{-2}$ density profile cannot persist all the way to the center of the collapsing cloud (Shu 1977; Mineshige \& Umemura 1997; Saigo \& Hanawa 1998).  After a time $t$, gas collecting at a rate $\dot M_0$ will dominate the potential out to a radius $r_1(t) \sim (t/t_0) r_0 $.  An $r^{-2}$ density distribution, with $M(r) \propto r$, is not globally self-gravitating inside $r_1$, and therefore cannot drive the bars within bars instability.  In nonrotating collapse (Shu 1997) the accumulated gas behaves like a point mass and the density at $r < r_1$ scales as $\rho \propto r^{-3/2}$, as in Bondi (1952) accretion. In inviscid collapse of a rotating fluid (Saigo \& Hanawa 1998), the centrifugal barrier forces the density to be roughly constant at small radii.  

Our model involves effective transport of angular momentum, and in this respect is closer to the case considered by Mineshige \& Umemura (1997). Assuming an $\alpha-$viscosity with fixed $\alpha$, they find that the surface density profile in the inner region steepens to $\Sigma \propto r^{-5/3}$, corresponding to a steep mean density distribution  $\rho\propto r^{-8/3}$.  The corresponding inflow rate scales as $\dot M \propto r^{1/3}$, and the inflow speed, $v\propto r$, falls far below the free-fall speed ($\propto r^{-1/3}$).  We also expect self-similar settling of gas at a mean speed $v(r) \sim r/t \ll v_0$ for $r \ll r_1$.  If we assume that angular momentum transfer is governed by global self-gravitational instabilities, instead of an $\alpha-$viscosity, then the gas must adopt a configuration where these instabilities are nearly quenched in order to transfer angular momentum on a timescale much longer than the dynamical time.  

The conditions for global instability depend on the details of the gravitational potential as well as the radial distributions of density, pressure, and angular momentum.  Because of the rapid cooling, gas pressure is negligible and we expect the system to remain strongly unstable at all radii where it is substantially self-gravitating (Shlosman et al. 1989).  The only way to suppress the instability, it seems, is for the density distribution to become sufficiently centrally concentrated that a large fraction of the gravitational potential at each $r$ is generated by the gas at much smaller radii.  This requires the mean surface density distribution to steepen to $\Sigma \propto r^{-2}$, corresponding to a mean density $\rho \propto r^{-3}$, so that there are roughly equal amounts of mass per decade of radius. Note that this is slightly steeper than the density distribution obtained by Mineshige \& Umemura (1997).  

The above argument assumes that angular momentum continues to be transported by global gravitational instabilities, {\it and} that fragmentation continues to be unimportant.  The latter assumption is much less secure at $r < r_1$ than it is further out.  If the gas remains isothermal at $\sim 10^4$ K, the Toomre $Q-$parameter would decrease $\propto r^{1/2}$ and the inflowing gas would form a thin disk.  Thus, fragmentation could seriously hamper the bars within bars instability at $r < r_1$.  If fragmentation is highly efficient, then the inflowing gas might simply lay down an isothermal stellar potential with a constant velocity dispersion $\sim v_0$.  Little gas would be deposited inside $r_1 (t)$.

It seems unlikely, however, that the outcome is this extreme.  Fragmentation should not deplete the gas density much below the threshold for {\it local} gravitational instability, $Q \sim 1$. Gas with the corresponding density, and a sound speed $\sim v_0$, would continue to accrete at a rate $\alpha \dot M_0$, where $\alpha$ is a viscosity parameter (Shlosman et al. 1990).  Even where fragmentation is suppressed, local nonaxisymmetric gravitational instabilities could continue to drive angular momentum transport with an effective $\alpha \sim O(1)$.  Other sources of angular momentum transport probably operate as well, such as turbulence stirred up by the fragmentation and star formation itself.  Therefore, although we are not able to predict exactly how much gas makes it all the way to the central region of the halo, we are probably justified in parameterizing the surviving mass flux as $\alpha \dot M_0$, with $\alpha$ assumed to be $\sim O(1)$. 

\section{Creation and growth of a ``quasistar"} 

Disk accretion persists as long as the infalling gas is able to radiate away the liberated binding energy. Given an accumulated mass of $M_* (t) \sim \alpha \dot M_0 t$ at $r \ll r_1$, we find that the luminosity generated outside a radius $r$ is given by
\begin{equation}
\label{Linner}
L(r,t)\sim  \alpha \dot M_0 {G M_* \over r} \sim \alpha^2 {v_0^5 \over G} \left[ {r \over r_1(t)} \right]^{-1}  .  
\end{equation}
Within a certain radius this radiation is trapped, the pressure builds up and the gas inflates into a pressure-supported cloud, which we dub a ``quasistar".  Since rotational support no longer dominates, we assume that the self-gravitational instabilities are finally quenched.  The condition for radiation trapping is given by 
\begin{equation}
\label{radtrap}
L(r,t) > L_{\rm Edd}(t) \left( 1 + {p_{\rm gas}\over p_{\rm rad}}\right)^{-1}   , 
\end{equation}
where $L_{\rm Edd} (t) \sim  4\pi \alpha c G\dot M_0 t/\kappa $ is the Eddington limit, given the appropriate opacity $\kappa$, for the accumulated mass $M_*(t)$.  $p_{\rm gas}$ and $p_{\rm rad}$ are the gas and radiation pressure, respectively, in the quasistar.  Once the quasistar mass exceeds a few solar masses (i.e., very early in its growth, since the mass is growing at $\sim 0.2 \alpha v_{10}^3 \ M_\odot$ yr$^{-1}$), the mean LTE radiation pressure exceeds the gas pressure; we will henceforth assume $p_{\rm rad} \gg p_{\rm gas}$.  The interior of the quasistar is hot enough to ionize hydrogen, allowing us to assume that the opacity is dominated by electron scattering.  Substituting $r_1(t) \sim (t/t_0) r_0$, we find that the radius of the quasistar is time-independent, and is given by 
\begin{equation}
\label{qrad}
r_* \sim  {\alpha \kappa v_0^3 \over 4 \pi G c} = 1.6 \times 10^{13} \alpha v_{10}^3 \ {\rm cm} . 
\end{equation}
Thus, the quasistar that is going to give rise to a supermassive black hole has a radius of 1 AU, for $\alpha v_{10}^3 \sim 1$.

Conditions in the interior of the quasistar are extremely sensitive to the mass inflow rate, which we express through its dependence on $\alpha$ and $v_0$. Denoting the quasistellar mass by $M_* = m_* \ M_\odot \propto t$, we find the mean density $\rho_* \sim 10^{-7} \alpha^{-3 }v_{10}^{-9} m_*$ g cm$^{-3}$,  mean pressure $p_* \sim 10^6 \alpha^{-4} v_{10}^{-12} m_*^2$ erg cm$^{-3}$, and mean temperature (in LTE) $T_*\sim 1.5 \times 10^5 \alpha^{-1}v_{10}^{-3} m_*^{1/2}$ K.  These estimates justify our assumptions that $p_{\rm rad} \gg p_{\rm gas}$ for  $m_* >$ a few, and that the opacity is dominated by electron scattering.

\section{Interior structure and evolution of the quasistar}

The characteristic specific entropy of the matter joining the quasistar increases with time, $s_* \equiv p_*/\rho_*^{4/3} \propto M_*^{2/3} \propto t^{2/3}$.  Since hydrostatic equilibrium demands $p \propto \rho^2 r^2 \propto \rho^{4/3} M(r)^{2/3}$ in the quasistellar interior, where $M(r)$ is the mass contained within a radius $r$, we conclude that each layer of matter added to the quasistar approximately conserves its specific entropy as the quasistar grows.  The positive entropy gradient, $s(r) \propto M(r)^{2/3}\propto r^{2/3}$, stabilizes the quasistar against convection, which would otherwise tend to homogenize the entropy.  This implies that each layer of the stellar interior is compressed by overlying material until the radiative diffusion time across the layer, $t_{\rm diff} (r)\sim \rho\kappa r^2 / c $, is roughly equal to the age of the quasistar. 

The interior structure of the quasistar is therefore characterized by a density profile $\rho \sim \rho_* (r/r_*)^{-2}$, pressure profile $p \sim p_* (r/r_*)^{-2}$, and temperature profile $T \sim T_* (r/ r_*)^{-1/2}$.   

These scalings apply as long as radiation pressure exceeds gas pressure.  However, the decreasing specific entropy toward the center implies that the ratio of radiation pressure to gas pressure decreases with decreasing $r$,
\begin{equation}
\label{gasradratio}
{p_{\rm rad} \over p_{\rm gas}}\sim  m_*^{1/2} \left( {r\over r_*}\right)^{1/2} , 
\end{equation}
implying that $p_{\rm gas} \sim p_{\rm rad}$ at small enough radii.  The gas pressure-dominated core comprises the earliest material laid down during the growth of the quasistar. It has a radius $r_c \sim r_* m_*^{-1} \sim 1.5 \times 10^{13} \alpha v_{10}^3 m_*^{-1}$ cm, temperature $T_c \sim 1.5 \times 10^5 \alpha^{-1}v_{10}^{-3} m_*$ K, and density $\rho_c \sim 10^{-7}\alpha^{-3} v_{10}^{-9} m_*^3$ g cm$^{-3}$.  The core mass is independent of $M_*$, $\alpha$, and $v_{10}$, and is roughly 1 $M_\odot$.   

Nuclear burning commences when the core temperature reaches $T_c \sim 10^6 - 10^7$ K, for a quasistar mass $m_* \sim (10 - 100) \alpha v_{10}^3$.  At this point the core density is not that dissimilar to densities inside main sequence stars, so the burning timescales are likely to be similar as well. The evolution timescale due to infall is much shorter, $t_{\rm ev} = M_* / \dot M_0 \sim 5 \alpha^{-1}v_{10}^{-3} m_*$ yr, so we do not expect nuclear burning to progress very far until the core temperature approaches $\sim 10^8$ K, for $m_* \sim 10^3 \alpha v_{10}^3$.  At this point the gravitational binding energy of the quasistar is $\sim 10^{13} \alpha^{-1} v_{10}^{-3} m_*$ erg g$^{-1}$ while the available nuclear binding energy is $\la 6 \times 10^{18} m_*^{-1}$ erg g$^{-1}$, if burning is confined to the core.  In order for the nuclear energy release to overpower the gravitational binding energy of the quasistar as a whole, the mass must satisfy $m_* < 700 \alpha^{1/2} v_{10}^{3/2}$.  Thus, by the time that nuclear reactions are able to run to completion, the available energy is probably insufficient to seriously affect the outer layers of the quasistar.  

It is the ultra-high infall rate, squeezing the core and raising its temperature beyond the thermostatic set points of thermonuclear reactions, that distinguishes the evolution of the quasistar from that of a normal Pop III star. At best, nuclear burning can provide a brief hiatus in the contraction of the core, which ultimately reaches temperatures $\ga 10^9$ K where neutrino losses become important.  

\section{Core collapse and initial growth of black hole} 

Continued compression by infalling matter prevents the core from losing energy radiatively and collapsing or becoming degenerate.  At sufficiently high temperatures, however, neutrino losses lead to core collapse and the formation of a black hole.  At $T_c \la 10^9$  K, the dominant neutrino loss mechanism is the URCA process, which is $\sim 300$ times faster than photo-neutrino production (Qian \& Woosley 1996; Dutta et al. 2004; Itoh et al. 1989; see also Koers \& Wijers 2005 for a summary of principal rates). At higher temperatures, pair annihilation becomes competitive with the URCA process, but because $\rho \propto T^3$ in the core, both mechanisms scale similarly with $T_c$.  Therefore, we may approximate the core cooling rate by $Q_c \sim 3 \times 10^{15} (T_c/ 10^9 \ {\rm K})^9$ erg s$^{-1}$ cm$^{-3}$. and the cooling timescale by $t_{\rm cool} \sim 4p_c/ Q_c $.  Setting this equal to $t_{\rm ev}$, we find that the core collapses when $m_* \sim 3600 \alpha v_{10}^3$ and $T_c \sim 5\times 10^8$ K. 

Details of the collapse depend on the angular momentum in the core as well as the precise core mass at the time of collapse.  If angular momentum is initially unimportant, the core should collapse at roughly constant temperature.  As the specific entropy decreases due to cooling, gas pressure begins to exceed radiation pressure and neutrino losses are dominated by the URCA process.  Because the core mass is rather low, collapse could get hung up by electron degeneracy pressure, but infalling matter from the quasistar envelope --- which continues to cool via neutrino losses --- would quickly drive the mass over the Chandrasekhar limit. Continued infall similarly circumvents neutron degeneracy, with the result that a black hole of a several solar masses forms in a few times the core free fall time.  

If the angular momentum of the core and surrounding material is too large to permit direct collapse to a black hole, then neutrino cooling will lead to a rotationally supported disk.  As material joins the disk, self-gravity will trigger a new round of ``bars within bars" instability, which will generate additional entropy (enhancing neutrino cooling) as well as facilitating collapse by removing angular momentum. 

The amount of matter that falls promptly into the black hole depends on the distribution of angular momentum in the $\rho \propto r^{-2}$ envelope of the quasistar.  The black hole will immediately swallow all the matter in the quasistellar envelope with a specific angular momentum $j = \Omega r^2 \la GM_{\rm BH}/c$. At one extreme, the specific angular momentum  at each radius could be a fixed fraction of the Keplerian value, implying $j (M) \propto M$ as a function of the enclosed mass $M(r)$.  In this case, the black hole does not grow immediately much beyond its initial mass.  At the opposite extreme, the internal redistribution of angular momentum within the quasistar could have led to solid-body rotation, implying that $j \propto M^2$.   In the latter case, the black hole mass could quickly swallow a fraction $\sim (GM_*/ r_* c^2)^{1/2}$ of the envelope's mass, assuming that the rotation rate reaches approximately the Keplerian value at $r_*$.  However, this amounts to only about $20 \alpha v_{10}^3 M_\odot$; therefore the prompt black hole mass is unlikely to exceed several tens of solar masses. 

\section{Subsequent black-hole growth}

After the initial collapse and prompt accretion phase, the growth of the black hole is regulated by angular momentum transport.  The  envelope continues to accumulate mass from infall, and the binding energy per unit mass increases with time, $v_* (t)^2 \sim G M_*(t)/ r_*$, where $r_*$ initially remains constant.  The black hole's gravitational sphere of influence has a radius $r_{\rm BH} \sim G M_{\rm BH} (t)/ v_* (t)^2 \sim (M_{\rm BH} / M_*) r_*$.  If angular momentum were unimportant, then the black hole would grow at the Bondi rate, $\dot M_{\rm Bondi } \sim v_*^3 /G$, and would swallow the quasistar in a free-fall time.  Thereafter it would grow at the infall rate, $\alpha \dot M_0$. 

Since angular momentum is important, a fraction of the binding energy released close to the black hole, $\epsilon \dot M_{\rm BH} c^2$, where $\epsilon \sim 0.1$ is the accretion efficiency, is transported outward by the torque.  If it is not radiated away from close to the hole --- by neutrino losses, since there is no free surface from which a wind can emerge --- this energy must react back on the inflow, slowing down the accretion. Let us assume that neutrino losses are negligible.  Initially, the region affected by the feedback energy is confined to the interior of the quasistar.  The total energy liberated by the time the black hole reaches mass $M_{\rm BH}$, $E_{\rm BH} \sim \epsilon M_{\rm BH} c^2$, affects the density profile inside the quasistar out to a radius $r_a \sim E_{\rm BH}/ (\rho_* r_*^2 v_*^2)$.   Note that the liberated energy is trapped inside the quasistar, rather than flowing through it in a quasi-steady state, because the radiative diffusion timescale at every radius is comparable to the age of the quasistar and the growth time of the black hole (\S~5).    

When $r_a$  reaches $r_*$, the liberated energy equals the binding energy of the quasistar and the latter begins to expand. This happens very early in the angular momentum-dominated growth phase, when the black hole mass has increased by an amount $\Delta M_{\rm BH} \sim  {GM_*^2 / \epsilon c^2 r_*} \la O(M_{\rm BH})$.   To show this, we estimate the rate at which the black hole swallows matter from the quasistar envelope.  The rate at which mass is {\it supplied} to the black hole's sphere of influence can be taken to be proportional to the Bondi rate, $\dot M_{\rm sup} \sim 4\pi \alpha_{\rm BH} \rho (r_{\rm BH}) v_*(t) r_{\rm BH }^2$, where $\rho (r_{\rm BH})$ is the density evaluated at the black hole radius of influence and $\alpha_{\rm BH} < 1$ is a parameter that describes the inefficiency of mass capture, e.g., due to a finite rate of angular momentum transport ($\alpha_{\rm BH}$ need not be the same as $\alpha$).  We expect the pressure and density distributions to be rather flat between $r_{BH}$ and $r_a$ because of the extra energy injection, and therefore use $v_*$ to estimate both $r_{\rm BH}$ and the free-fall speed at the radius of influence.  We also take $\rho (r_{\rm BH}) \sim \rho (r_a) \sim \rho_* (r_*/r_a)^2$.  Note, however, that even if the density increases $\propto r^{-\beta}$ at $r < r_a$, our prescription gives a lower limit to $\dot M_{\rm sup}$ provided that $\beta  < 14/5$.  

The rate at which matter actually reaches the black hole is suppressed by a further factor, due to the back reaction of the energy flux inside the radius of influence (Gruzinov 1998; Blandford \& Begelman 1999; Narayan, Igumenshchev \& Abramowicz 2000; Quataert \& Gruzinov 2000).  In the absence of a wind that removes energy and/or angular momentum, the accretion rate is reduced to $\dot M_{\rm BH} \sim \epsilon^{-1} (v_*/c)^2 \dot M_{\rm sup}$ (Blandford \& Begelman 1999, 2004).  We then find that the accretion rate is
\begin{equation}
\label{maccnonu}
\dot M_{\rm BH} \sim 3 {\alpha_{\rm BH} \over \epsilon^3}  {c^3\over G } \left({v_* \over c} \right)^9.
\end{equation}
Since $v_* \propto M_*^{1/2}$, $\dot M_{\rm BH} \propto M_*^{9/2}$ is a steeply increasing function of $M_*$.  Comparing it to the inflow rate onto the quasistar, we find that the feedback energy equals the binding energy of the quasistar before the black hole mass has doubled. Thus Bondi accretion, even modified by feedback and a finite rate of angular momentum transport, should quickly bring the quasistar to the point where its evolution is driven by feedback from the black hole.  

The feedback flux does not blow apart the quasistar, since this would stop the growth of the black hole and therefore the feedback.  Instead the quasistar expands gradually, allowing the black hole accretion rate to adjust so that the feedback energy flux equals the Eddington limit for the instantaneous quasistar mass, $\dot M_{\rm BH} \sim 2 \times 10^{-3} (\epsilon/0.1)^{-1}(m_*/10^5) M_\odot$ yr$^{-1}$. The feedback energy flux exceeds the Eddington limit for the black hole by a factor $M_*/M_{\rm BH}$; thus, the black hole grows at a super-Eddington rate as long as $M_* > M_{\rm BH}$.  This configuration requires most of the feedback flux to be carried by convective motions inside the quasistar, since the enclosed mass at $r_{\rm BH} < r < r_*$ is a steeply increasing function of radius.  However, one can show that the required convective velocity, while larger than the mean inflow speed, is much smaller than the local free fall speed at all $r$. If the quasistar mass continues to increase at the constant rate $\alpha \dot M_0$, then the black hole mass evolves according to 
\begin{equation}
\label{BHmass2} 
M_{\rm BH} (t) \sim  4 \times 10^5 \alpha v_{10}^3 \left( \epsilon \over 0.1 \right)^{-1} \left({t \over 10^7 \ {\rm yr}}\right)^2  M_\odot ,
\end{equation}
i.e., $M_{\rm BH} \propto M_*^2$.  

To determine the evolution of the quasistar's structure in response to feedback, we estimate $\dot M_{\rm BH}$ using the modified Bondi rate discussed above.  If we assume the density to be roughly uniform within the quasistar (outside $r_{\rm BH}$), we have
\begin{equation}
\label{maccnonu2}
\dot M_{\rm BH} \sim 3 \alpha_{\rm BH}   {c^3\over \epsilon G } \left({M_{\rm BH} \over M_*} \right)^2\left({v_* \over c} \right)^4.
\end{equation}
Equating this to the Eddington-limited rate (assuming electron scattering opacity) and using eq.~(\ref{BHmass2}) and the assumed infall rate onto the quasistar, we obtain
\begin{equation}
\label{revol}
r_* \sim 2\times 10^{15} \alpha^{-1} v_{10}^{-3} \left({\alpha_{\rm BH}\over 0.01}\right)^{1/2}\left( \epsilon \over 0.1 \right)^{-1} \left({m_* \over 10^5} \right)^{3/2}  \ {\rm cm} .
\end{equation}

Neutrino losses are rapidly quenched by the expansion of the quasistar.  Radiation pressure dominates throughout the envelope, and the temperature (in LTE) decreases linearly with time (and with $M_*$), $T_* \sim 4 \times 10^5 (\alpha_{\rm BH}/0.01)^{-1/2} \alpha v_{10}^3(\epsilon/ 0.1) (m_*/10^5)^{-1}$ K. Within the black hole's radius of influence, the pressure varies $\propto r^{-3/2}$ (not $\propto r^{-5/2}$, as in ordinary Bondi accretion, because of the feedback), and $T\propto r^{-3/8}$ can exceed $T_*$ by a factor as large as $(\epsilon c^2 / v_*^2) \sim 40 (\alpha_{\rm BH}/0.01)^{3/16} (\alpha v_{10}^3)^{-3/8}(m_*/10^5)^{3/16}$, close to the black hole.  Such temperatures are inadequate to produce a significant neutrino flux when the black hole grows much beyond its initial mass.  

The effective temperature of the quasistar's photosphere is also expected to decrease, implying that the quasistar is unlikely to be a significant source of hard UV radiation when it has grown beyond $\sim 10^4 M_\odot$.  Up to this point, the rate of production of ionizing photons is very high, of order $10^{55} {\rm photons \ s^{-1}}$; but since this phase lasts for $\la 10^5{\rm yr}$, the total output falls far short of the requirement for reionizing the Universe.  Similarly, the quasistar produces $\simeq 10^{50} {\rm photons \ s^{-1}}$ in the Lyman-Werner band, but can keep the molecular hydrogen in its surroundings photodissociated only for $\la 10^5{\rm yr}$. These estimates correspond to a spherical photosphere at $r_*$, but we note that photospheric temperatures could be even lower if the photosphere is strongly flattened by rotation.

The above estimates are valid only as long as $T_* \ga 10^4$ K, corresponding to 
\begin{equation}
\label{mstarmax}
M_* < 4\times 10^6 \alpha v_{10}^3 \left({\alpha_{\rm BH}\over 0.01}\right)^{-1/2}\left( \epsilon \over 0.1 \right)  \ M_\odot 
\end{equation}
and
\begin{equation}
\label{mmax}
M_{\rm BH} <  9\times 10^5 \alpha v_{10}^3 \left({\alpha_{\rm BH}\over 0.01}\right)^{-1} \left( \epsilon \over 0.1 \right)    \ M_\odot .
\end{equation}
For the fiducial parameters, the black hole mass at this stage is almost as large as that of the quasistar; further growth can occur at the Eddington limit of the black hole.  However, we emphasize the uncertainty in parameters such as $\alpha_{\rm BH}$ and $\alpha$.  (We use different fiducial estimates of $\alpha$ and $\alpha_{\rm BH}$ because the latter represents viscous transport of angular momentum while the former represents a removal of gas from the inflow due to fragmentation and star formation.) 

At lower temperatures, the Planck mean opacity (which is relevant for calculating the radiation force in LTE, and therefore the Eddington limit) becomes very sensitive to temperature (Mayer \& Duschl 2005), increasing sharply at $T_* \la 10^4$ K and then decreasing to several orders of magnitude below the electron scattering opacity as the temperature declines further.  The sharp decrease in opacity would affect the photosphere at an even earlier stage in the quasistar's evolution. 

Finally, we note the existence of an alternative evolutionary scenario in which $\alpha_{\rm BH}$ is so small that feedback does not regulate the structure of the quasistar.  If $\alpha_{\rm BH}$ is essentially zero (in practice, $\ll 10^{-6}$ when $m_* \sim 10^5$), then the centrifugal barrier forms a wall within the quasistar at $r \sim r_{\rm BH}$.  $r_*$ is once again constant and the  temperature at $r_{\rm BH}$ scales as $M_* /M_{\rm BH}^{1/2}$.  If this ratio increases with time, following the initial collapse and prompt accretion phase, then neutrino losses remain important within the black hole's sphere of influence.  We are then justified in assuming that self-gravitational instabilities transport angular momentum effectively.  Provided that nearly all of the liberated binding energy is carried away by neutrinos, we deduce that the black hole grows at such a rate that $M_{\rm BH} \propto M_*^2 \propto t^2$.   Adopting $T(r_{\rm BH}) = 10^9 T_9$ K as the threshold for rapid neutrino cooling, we find that the black hole mass (in solar units) grows according to 
\begin{equation}
\label{BHmass} 
m_{\rm BH} \sim 225 \alpha^{-2} v_{10}^{-6} T_9^{-2} \left({m_* \over 10^5}\right)^2 \sim 9\times 10^4  T_9^{-2} \left({t \over 10^7 \ {\rm yr}}\right)^2 
\end{equation}
This rate is not much smaller than the Eddington-limited rate, eq.~(\ref{BHmass2}), in the presence of feedback.  Given the convergence of these two extreme estimates, we are reasonably confident that rapid black hole growth up to masses $\sim 10^4 - 10^6 M_\odot$ is possible under the conditions postulated here. 

The discussion in \S\S~3--7 can be generalized to halos with $T_{\rm vir} < 10^4$ K.  If molecular hydrogen cools the gas down to $\sim 200$ K, then runaway collapse without fragmentation could occur in halos with correspondingly low virial temperatures.  The characteristic infall speed is then $v_{10} \la 0.2$, implying inflow rates of a few thousandths of a solar mass per year.  Nevertheless, a quasistar with a hot dense core will eventually develop, and will ultimately collapse to form a black hole due to runaway neutrino cooling.  The mass of the prompt black hole is insensitive to $v_{10}$, and therefore is still $\la 20 M_\odot$.  However, the mass of the quasistar at this stage scales as $v_{10}^3$; thus the quasistar mass is only a few times that of the black hole.  Moreover, the black hole could not reach more than a few thousand solar masses before the growth rate becomes sub-Eddington, and limited by the inflow of gas into the quasistar.

\section{Evolution of the black hole population}

How large a population of black holes is likely to result from gravitational instability of gas discs in high redshift halos?  Given the threshold of $T_{\rm vir} > 10^4$ K for efficient cooling, and therefore for ``bars  within bars'' instability, we can trace the co-evolution of the black holes and their hosts.  A $T_{\rm vir} \sim 10^4$ K halo has a mass between $10^7M_\odot$ and $10^9M_\odot$ at redshift $6<z<20$.  The black hole forming in such a host could grow at the super-Eddington rate given by eq.~(\ref{BHmass2}) until it reaches $\sim 10^6 M_\odot$, at which point its mass would approach that of the quasistar.  Thereafter it  could grow, by Eddington-limited accretion (or the infall rate, if smaller), to an even larger mass.  However, the growth of the black hole can also be terminated earlier by lack of fuel, i.e., by using up all the available gas in the unstable disc (if $f_{d}$ is much smaller than unity) and/or if star formation depletes the inflow.  Therefore, we do not assume that black holes grow rapidly to the maximum allowed mass. Indeed, we will see below that our mechanism can provide the seeds for all present-day SMBHs, even if its efficiency is quite low.  

More massive halos, with $T_{\rm vir} \gg 10^4$K, are probably prone to fragmentation and star formation, which would inhibit instability and therefore the formation of a black hole by this process.  Using the Press--Schechter formalism (Sheth \& Tormen 1999), we estimate that the number density of halos with virial temperature $T_{\rm vir} > 2\times 10^4$ K  ($T_{\rm vir} > 5\times 10^4$ K) is about $10\%$ ($1\%$) of the density of halos with $10^4$ K $<T_{\rm vir} < 2\times 10^4$ K ($10^4$ K $<T_{\rm vir} < 5\times 10^4$ K). More massive halos make an even smaller contribution.   Since the contribution of halos with $T_{\rm vir} \gg 10^4$  K is negligible, we can estimate the black hole density by integrating over all host halos with $T_{\rm vir} > 10^4$ K. 

\begin{figure}
\includegraphics[width=8cm, height=8cm]{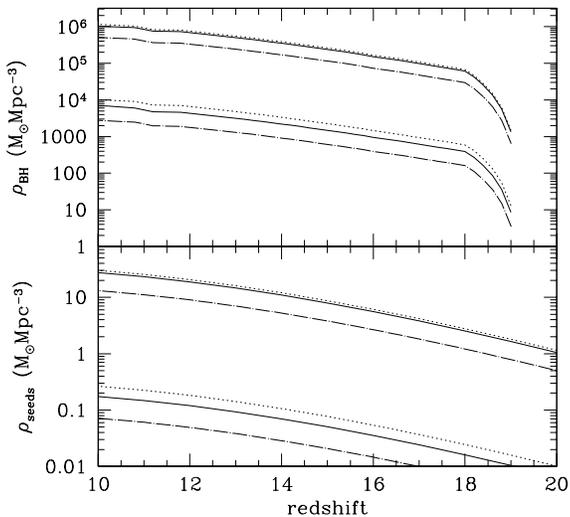}
\caption{Lower panel: comoving density of $20 M_\odot$ black hole seeds as a function of redshift. Upper panel: comoving density of black holes, assuming continuous formation and growth of seeds to $10^6 M_\odot$ according to eq.~(\ref{BHmass2}), starting from $z=20$.   Solid line: Mestel disc, dotted line: rigid disc, dot-dashed line: exponential disc. The upper set of lines assumes $f_{d}=0.5$, the lower set assumes $f_{d}=0.1$. }
\label{fig:BHdensity}
\end{figure}

Among all halos with $T_{\rm vir}> 10^4$ K, only those with a low enough spin parameter (given $f_d$) will host a disc unstable to ``bars within bars" instability.  Assuming a seed black hole mass of $20 M_\odot$, we plot the comoving mass density of black hole seeds in Fig.~\ref{fig:BHdensity}. The mass density is small, but this process can nevertheless seed most of the systems which will evolve into the local galaxies where SMBHs have been found. For a given local halo, the Extended Press--Schechter formalism can be used to estimate the average number of progenitors with $T_{\rm vir} \sim 10^4$ K at $z>10$. The probability of black hole formation depends also on the amount of gas which condenses to form a disc (see \S~2). If the fraction of gas typically ending up in the disc is $f_d\approx 0.5$, we find that this process needs to operate only until $z\simeq 18$ in order to supply seeds to all present-day halos of Milky-Way, or larger, size. If black hole formation proceeded to $z\simeq 14$, then all halos with mass $> 10^{11} M_\odot$ today could have been seeded (see Barth, Greene \& Ho 2005). If $f_d$ was lower, the black hole formation process would have had to continue for longer, in order to form seeds in the progenitors of most local galaxies.  These constraints would be eased by an early generation of black hole seeds, formed in a small fraction of halos with $T_{\rm vir} \ga 200$ K, before H$_2$ is photodissociated by the first stars. 

The process of seed formation can be widespread enough to account for subsequent SMBH evolution, even if the subsequent black hole growth (\S~7) is inefficient in most high-redshift halos. We note, in fact, that black holes with masses well below $\sim 10^6 M_\odot$ are expected in local faint AGN (Barth, Greene \& Ho 2005). The growth of black hole seeds up to  $\sim 10^4 - 10^6 M_\odot$ cannot therefore happen in all high-redshift systems. 

We can estimate the upper limit to the total black hole mass density predicted by our model by assuming continuous formation of seed black holes, and adopting eq.~(\ref{BHmass2}) to estimate their early growth.  We let black holes grow until either they have consumed all the gas, or the infall of matter onto the black hole greatly decreases, at  $M_{\rm BH} \approx 10^6 M_\odot$. The comoving density of halos is estimated using the Sheth \& Tormen (1999) halo mass function. The black hole mass density shown in Fig.~\ref{fig:BHdensity} must be compared to the density that we observe at low $z$, i.e., $\rho_{\rm BH}(z=0)\approx 3-5 \times 10^5 M_\odot \ {\rm Mpc^{-3}}$ (Yu \& Tremaine 2002; Aller \& Richstone 2002; Marconi et al. 2004; Fabian \& Iwasawa 1999; Elvis, Risaliti \& Zamorani 2002) and $\rho_{\rm BH}(z=3)\approx 4-5 \times 10^4 M_\odot \ {\rm Mpc^{-3}}$ (Merloni 2004). At $z=6$, a lower limit to the  black hole density is obtained by  integrating the observed luminosity function, $\rho_{BH}(z=6)\approx {\rm few} \times M_\odot  \ {\rm Mpc^{-3}}$ (Fan et al. 2004). This density, however, includes only black holes with masses $\ga 10^9 M_\odot$. To obtain a rough estimate of the total black hole density at $z=6$, we adopt So\l tan's (1982) argument.  We extrapolate the luminosity function up to $z=6$, assuming the redshift dependence given by Richards et al. (2005), integrate the emitted quasar luminosity, and convert it into a black hole density by normalizing to the $z=0$ SMBH density. We do not correct here for obscured quasars, or assume a radiative efficiency, and therefore consider the density that we find, $\rho_{BH}(z=6)\approx 1-3 \times 10^3 M_\odot \ {\rm Mpc^{-3}}$, more of an indication than a robust estimate. 

The densities shown in Fig.~\ref{fig:BHdensity} must be regarded as upper limits to the black hole density, as we have not included any effect that can interrupt or disturb black hole formation and growth. We discuss in the next section how the hierarchical framework for structure formation can modify this simple picture. We also recall that, eventually, efficient star formation occurs in these halos, competing for the gas supply and possibly limiting the mass available for black hole accretion.

\section{Discussion and conclusions}

We have presented a scenario for the accumulation of gas in the centers of dark matter halos with $T_{\rm vir} \ga 10^4$ K, the initial collapse of the gas to form a seed black hole, and the subsequent early growth of a supermassive black hole.  This mechanism can lead naturally to the super-Eddington growth of black holes up to masses $\sim 10^6 M_\odot$, as early as redshifts $10-20$. Given additional growth to $\sim 10^9 M_\odot$ at close to the Eddington rate, the model can account for the population of quasars observed at $z\sim 6$ (Fan et al. 2004).  Even without significant growth after the formation phase, this mechanism could produce the seeds for all supermassive black holes inferred to exist in the local universe. 

We argue that global self-gravity triggers the ``bars within bars" instability (Shlosman et al. 1989, 1990), under certain conditions, as gas forms a rotationally supported thick disk in the center of the halo.  On scales much smaller than the disc radius, and times shorter than the free-fall time, quasi-steady inflow is a better representation of the infall than a monolithic collapse. Local, or quasi-local sources of ``viscosity", such as those due to magnetic fields, turbulence, or radiation drag, are not required to transport the angular momentum that inhibits black hole formation.  In metal-free halos with little molecular hydrogen, this behavior is possible once the virial temperature exceeds $\sim 10^4$ K (Oh \& Haiman 2002).  Under these conditions, gravitational instabilities can transport angular momentum effectively from scales of several kiloparsecs down to scales initially as small as $\sim 1$ AU, at a fraction of a solar mass per year (for a characteristic infall speed $v_0 \sim 10$ km s$^{-1}$).  We suggest that inflow is most efficient in halos where $T_{\rm vir}$ does not exceed a few times $10^4$ K, since fragmentation of the infalling gas is unlikely to be efficient in this case (Bromm \& Loeb 2003). As the mass in the center builds up, global instabilities may be quenched in the inner regions, but local gravitational instabilities could continue to drive a substantial inflow, even if a certain amount of star formation occurs.  

Instability occurs only where the halo spin parameter, $\lambda_{\rm \rm spin}$, falls below a threshold value that depends on $f_d$, the fraction of gas that forms the disc.  For $f_d \ga 0.5$ the threshold is comparable to the mean spin parameter predicted by simulations, and $> 20$\% of all $T_{\rm vir} \sim 10^4$ K halos should exhibit instability.  Even a value of $f_d$ as small as $0.1$ leads to instability in $\ga 1$\% of halos and a significant seed population of black holes. 

In halos with the low angular momentum required to trigger black hole formation, the rapid infall of gas leads to a mass accumulation much larger than that expected in a mini-halo with an average spin parameter. The formation of a ``standard'' Pop III star  (Bromm, Coppi \& Larson 1999, 2002; Abel, Bryan \& Norman 2000) is therefore suppressed in favor of a massive ``quasistar" (\S~4).  We suggest that the much smaller mini-halos (with virial temperature well below $10^4$ K) that form the first stars are distinct from the halos that form the seeds of supermassive black holes, although the former may be precursors of the latter.  Photodissociation of molecular hydrogen, possibly by a small population of Pop III stars, would suppress fragmentation of the infalling gas.  It is therefore possible that the formation of seed black holes follows an earlier epoch of star formation, as the ``quasistar'' itself is not a significant source of photo-dissociating photons for long.  The epoch of black hole formation must happen early enough, however, that the Universe is not highly metal enriched --- later episodes of star formation would enrich the environment of seed black holes.

The most important conclusion of our model is that the ``quasistar" formed by the accumulating gas has a low-entropy, gas pressure-dominated core surrounded by a radiation pressure-dominated envelope.  As matter piles onto the quasistar, the core is squeezed until its temperature approaches $10^9$ K (typically when the envelope mass reaches a few thousand $M_\odot$).  Cooling by thermal neutrinos then leads to core collapse and the formation of a seed back hole of $\sim 10-20 M_\odot$.  This is a novel application of neutrino cooling, which has been invoked previously in connection with hyperaccretion onto neutron stars in supernovae (Colgate 1971; Chevalier 1989; Houck \& Chevalier 1991) and common envelope binaries (Chevalier 1993; Brown, Lee \& Bethe 2000), and onto black holes in gamma-ray bursts (Narayan, Paczy\'nski \& Piran 1992; Woosley 1993; Popham, Woosley \& Fryer 1999; Narayan, Piran \& Kumar 2001).  It is difficult to set up the necessary conditions for efficient neutrino cooling, since radiation pressure generally prevents the accretion rate from reaching the required level from below, unless the viscosity parameter $\alpha$ is extremely small (Chevalier 1996).  Previous models have circumvented this problem by invoking strong radiation trapping in a steady-state (or slowly varying) accretion flow.  In our case the inflow sets up favorable conditions in a time-dependent fashion by establishing a steep positive entropy gradient in the quasistar, with only mild radiation trapping.

The subsequent evolution of the black hole can be very fast, with growth to more than a million solar masses possible in less than a Salpeter timescale.  Even taking account of strong energy feedback driven by angular momentum transport, we conclude that black holes can accrete at the Eddington rate {\it for the quasistar mass}, which exceeds the Eddington rate for the black hole by a factor $M_*/ M_{\rm BH}$.   For steady infall onto the quasistar, this corresponds to a black hole mass increasing with time as $M_{\rm BH} \propto t^2$.  

If black hole growth (eq.~\ref{BHmass2}) proceeded undisturbed in all halos satisfying the instability criterion with $T_{\rm vir}> 10^4$ K, then the total mass density in supermassive black holes would become comparable to the local one already at high redshift.  However, a number of effects can limit this initial phase of rapid growth.  Limitations intrinsic to the halo include the overall mass supply that participates in the infall, as well as removal of matter from the inflow by star formation.   Moreover, the halos and their embedded black holes do not grow in isolation.  Halos susceptible to the ``bars within bars" instability represent high peaks in the field of density fluctuations (Tegmark et al. 1997; Madau et al. 2004). Therefore, they should experience an enhanced number of major mergers with respect to ``average" halos at the same redshift.  Halo major mergers can modify our basic results in two ways.  First, cosmological simulations show that the spin parameter of a halo typically increases after a major merger (Vitvitska et al. 2002). On the other hand, the spin parameter decreases after a long series of minor mergers.  Major mergers, therefore, should delay the triggering of instabilities, at least until a sufficient number of minor mergers has lowered the spin parameter again.  Second, a major merger could destroy the coherence of the ``bars within bars" process. By interfering with the infall of matter onto the quasistar, a violent encounter could hasten the depletion of the mass supply well before the upper limits discussed above are reached.  Such a disturbance is unlikely to modify the interior structure of the existing quasistar or suddenly stop the growth of the black hole, however, since these involve only the very core of the system.

Seeding of larger halos by black hole mergers could also be limited by the ``gravitational rocket" effect, the recoil due to the net linear momentum carried away by gravitational waves in the coalescence of two black holes (Madau et al. 2004, Haiman 2004, Yoo \& Miralda-Escud\'e 2004).  The recoil velocity still has large uncertainties, but can easily exceed $\sim 100 \ \rm{km \ s^{-1}}$, comparable to the escape velocity from shallow halo potentials. Volonteri \& Rees (2006) estimate that up to $50\%$ of black holes merging in high-redshift halos can be ejected due to the gravitational rocket effect. 

Despite these potential sources of inefficiency, the mechanism we have outlined could be the principal route leading to supermassive black hole formation in galactic nuclei.  The main elements of the model --- particularly the cascading infall via ``bars within bars" instability, and the formation and evolution of the quasistar with runaway neutrino cooling --- should be testable via numerical simulations.  We hope that such simulations can be undertaken shortly.

\section*{Acknowledgments}

This work was supported in part by NASA Beyond Einstein Foundation Science grant NNG05G192G, NSF grant AST-0307502, and the University of Colorado Council on Research and Creative Work. MCB thanks the Institute of Astronomy and the Master and Fellows of Trinity College, Cambridge, for their hospitality.

\end{document}